\documentclass{article}
\begin{document}
\newcommand{\be}{\begin{equation}}
\newcommand{\p}{\partial}
\newcommand{\nn}{\nonumber}

\newcommand{\ee}{\end{equation}}
\newcommand{\beqn}{\begin{eqnarray}}
\newcommand{\eeqn}{\end{eqnarray}}
\newcommand{\f}{\frac}
\newcommand{\w}{\wedge}

\begin{titlepage}
\flushright{WIS/9/05-MAY-DPP}
\flushright{hep-th/0505039}

\vspace{1in}

\begin{center}
\Large
{\bf Branes in $L^{(p,q,r)}$ }
\vspace{1in}

\normalsize

\large{ D. Gepner and Shesansu Sekhar  Pal }\\

\normalsize
\vspace{.7in}

{\em Weizmann institute of  science,\\
76100 Rehovot, Israel }\\
{\sf doron.gepner, shesansu.pal@weizmann.ac.il}
\end{center}

\vspace{1in}

\baselineskip=24pt

\begin{abstract}

We have found the solution to the back reaction of putting a stack of 
coincident D3 and D5 branes in $R^{3,1}\times M_6$, where $M_6$ is 
constructed from 
an infinite class of Sasaki-Einstein spaces, $L^{(p,q,r)}$. The  
non-zero fluxes associated to 2-form potential 
 suggests the presence of a non-contractible 2-cycle in this 
geometry. The radial part  of the warp factor has the usual form and possess 
the  cascading feature. We argue that generically the duals of these SE 
spaces will have irrational central charges.  


\end{abstract}

\end{titlepage}

\section{Introduction}
The construction of various 5-dimensional compact manifold is of great 
importance in string theory especially in the study of  field theory. 
Wrapping D-branes along the cycles of  the Calabi-Yau constructed from 
this along with adding 
some extra non-compact directions, i.e. putting  stacks of coincident 
 D5 branes and D3 branes branes along both the compact 
and non-compact directions   teaches us a lot  about the corresponding dual
field theory.  The field theoretic observables are possible to calculate from 
the gravitational quantities due to a masterly duality between these 
theories, i.e. ADS/CFT correspondence \cite{jm}, for a review \cite{agmoo}. 
Otherwise, the computation of
the field theoretic quantities like scaling dimension or R-charge of operators  and central charges would have been very difficult to do. \\

Some of the examples, which preserve some non-zero supersymmetry, that have 
been studied so far are $S^5, T^{(1,1)}\cite{co,kw,kt}, 
Y^{(p,q)}\cite{gmsw1,gmsw,bfhms}$\footnote{p,q and r are coprime integers.}, and deformation thereof \cite{ks,pt}. These manifolds 
fall into a class called 
Sasaki-Einstein(SE) type i.e. it admit a Killing spinor. 
Of-course, the last one was
the most general until very recently \cite{clpp}, where an infinite set of 
 completely new non-singular Sasaki-Einstein manifolds with explicit form 
of the metric is given, which has been dubbed as $L^{(p,q,r)}$, for some choice of $p,q,r$. $Y^{(p,q)}$ 
spaces can be obtained from $L^{(p,q,r)}$ as $Y^{(p,q)}=L^{(p-q,p+q,p)}$ 
\cite{clpp}. 
These SE geometries have been obtained by euclideanising along with taking a 
special limit called ``BPS scaling limit'' of the five dimensional 
rotating Kerr-ADS black hole solution given in \cite{hht}. This way of 
generating SE geometries was started out with \cite{hsy}, where these authors
were able to generate the $Y^{(p,q)}$ spaces from euclideanising the same five 
dimensional Kerr-ADS geometry and taking another special limit. 
It is interesting to note that the $Y^{(p,q)}$ and $L^{(p,q,r)}$ spaces are 
obtained by taking 
different limits of the same two parameter family of five-dimensional 
rotating blackhole solution of \cite{hht}.  

The isometry of  $L^{(p,q,r)}$ is $U(1)\times U(1)\times U(1)$ and becomes a 
smooth non-singular space for $q\ge p> 0$ and $p+q >r >0$, but the Calabi-Yau that we shall construct has a singularity at $r=0$. This isometry goes 
over to $SU(2)\times U(1)\times U(1)$ for $p+q=2r$ \cite{clpp}. The $U(1)^3$
isometry is described by certain linear combinations of Killing
vectors: $\f{\p}{\p\tau}, \f{\p}{\p \psi}$ and $\f{\p}{\p \phi}$. It is 
the orbit of the first Killing vector which determine whether the SE space 
will be a quasi-regular or regular \cite{clpp}. The former and the later 
corresponds to 
the closed and open orbits of the $\f{\p}{\p\tau}$ Killing vector respectively.
Its not clear 
whether this space has the topology of $S^2 \times S^3$ \footnote{This is 
being confirmed to us by Chris Pope that it admits the topology of 
$S^2\times S^3.$ }. In any case , 
we  argue
that this geometry do admit a   non-contractible two cycle by 
turning on  NSNS fluxes and finding the situation where the periods 
become non-vanishing, implies the presence of a 2-cycle and 
support pope's claim of the topology of this space to be $S^2\times S^3$.

Even though, we know the explicit form of the metric 
obtained through this way, as mentioned above, still it would be interesting 
to derive it 
from 11-d supergravity following \cite{gmsw} and to see whether this falls in 
the class studied there or it  lead to a new way of classifying various 
supergravity solutions. 

In this paper we would like to find solutions to Type IIB supergravity 
constructed on this SE spaces by constructing a cone on this space and 
putting stacks of D5 branes on the apex of cone along with a stack a D3 branes
which are extended along the flat four directions. As  expected, the 
solution should exhibit the cascade feature and which the solution does. 

\section{The $L^{(p,q,r)}$ geometry}
 
The five dimensional $L^{(p,q,r)}$ geometry  is described by the following 
form of the metric, which we review following \cite{clpp}
\beqn
\label{lpqr}
ds^2_5&=&\f{1}{\lambda}\bigg[ (d\tau+\sigma)^2+\f{\rho^2}{4\Delta_x}dx^2+\f{\rho^2}{\Delta_{\theta}}d\theta^2+\f{\Delta_x}{\rho^2}\bigg(\f{sin^2\theta}{\alpha}d\phi+\f{cos^2\theta}{\beta}d\psi \bigg)^2+\nn \\& &\f{\Delta_{\theta} sin^22\theta}{4 \rho^2}\bigg(\f{\alpha-x}{\alpha}d\phi-\f{\beta-x}{\beta}d\psi \bigg)^2\bigg],
\label{lpqr}
\eeqn 
where $\sigma$ is a one form and the form of it and $\Delta_x, \Delta_{\theta}$
and $\rho^2$ are given by
\beqn
& & \sigma=\f{(\alpha-x)sin^2\theta}{\alpha}d\phi+\f{(\beta-x)cos^2\theta}{\beta}d\psi; \quad \rho^2=\Delta_{\theta}-x\nn \\& &
\Delta_x=x(\alpha-x)(\beta-x)-\mu; \quad
\Delta_{\theta}=\alpha cos^2\theta+\beta sin^2\theta.
\eeqn

The ranges of the coordinates are $0< \theta <\pi/2, x_1 < x < x_2, 0 <\phi, 
\psi <2 \pi$, where $x_1$ and $x_2$ are the two lowest roots of $\Delta_x=0$ 
equation and $\tau$ is a compact coordinate. The parameter $\mu$ can be set to any constant value by rescaling 
$\alpha, \beta, x$. Since, we are not going to provide either the exact form of
$x_1$ and $x_2$ or the period of $\tau$, so we shall stay with $\mu$. In any 
case, $\mu$ 
do not appear in $x_1,x_2$. For $\mu=0$, one gets the metric
of $S^5$ \cite{clpp}. However, we do write down the properties of the roots of
$x_i$ with $i=1,2,3$, which might be helpful for the computation of $x_1,x_2$.
\beqn
& &\sum^3_{i=1}x_i=\alpha+\beta;\quad \prod_{i<j} x_i x_j=\alpha\beta; \quad x_1x_2x_3=\mu,\nn \\& &
x_1+x_2=\f{1}{\alpha\beta}[x^2_1+x^2_2+x_1x_2+\alpha\beta].
\eeqn 
A relation between $x_1-x_2$ will help us to obtain the exact form of 
$x_1,x_2$. Even if  we do not know that still we can proceed by 
assuming that 
$x_2-x_1\equiv\Lambda$, with $\Lambda$ a real number as $x_1, x_2$ are real 
numbers. Then the two roots are
\beqn
x_1&=&\f{1}{6}[2\alpha\beta-3\Lambda-\sqrt{-12\alpha\beta+4\alpha^2\beta^2-3\Lambda^2}],\nn \\
x_2&=&\f{1}{6}[2\alpha\beta+3\Lambda-\sqrt{-12\alpha\beta+4\alpha^2\beta^2-3\Lambda^2}].
\eeqn

From this there follows a restriction on $\alpha,\beta$ and $\Lambda$ 
that is $4\alpha^2\beta^2 > 12\alpha\beta+3\Lambda^2$. Which then imply that 
the volume of this SE paces are irrational like those found in $Y^{(p,q)}$
case and by ADS/CFT this implies that the central charges associated to 
the corresponding field theory is also   irrational. Of course, for  a
special case like 
\be
\label{rational_roots}
4\alpha^2\beta^2 = 12\alpha\beta+3\Lambda^2,
\ee
 one will have rational roots.


The Calabi-Yau that we are interested in is 
\be
\label{cy}
ds^2_6=dr^2+r^2 ds^2_5,
\ee
and the volume of the 5-d geometry $L^{(p,q,r)}$ is given by
\be
Vol_5=\f{\Delta\tau\pi^2}{4\alpha\beta\lambda^\f{5}{2}} (x_2-x_1)
(\alpha+\beta-x_1-x_2),
\ee 
where $\Delta\tau$ is the period of $\tau$ and $x_3=\alpha+\beta-x_1-x_2$. 
For completeness, let us mention the form of it and is given by
\be
x_3=\f{2}{6}[3(\alpha+\beta)-2\alpha\beta+
\sqrt{4\alpha^2\beta^2-12\alpha\beta-3\Lambda^2}].
\ee
Using the form of $x_3$ and $x_2-x_1=\Lambda$ 
 yields the volume of 5-d SE space as
\be
Vol_5=\f{\Delta\tau\pi^2\Lambda}{4\alpha\beta\lambda^\f{5}{2}} ~~x_3.
\ee
Now if the period of $\tau$ is anything other then $\f{2\pi}{x_3}$ then 
the volume is irrational and hence the central charge, C, of the corresponding
dual theory, which is $C\sim \f{1}{Vol_5}$ also irrational. However, if 
the period is exactly $1/x_3$ then the volume and central charge are
rational. If we assume that the volume (central charge) of $L^{(p,q,r)}$
spaces are rational then its highly unlike that for some combination 
of $p,q,r$ one will generate irrational volume (central charges), as 
$Y^{(p,q)}$ are special cases of $L^{(p,q,r)}$. However, the converse
looks more plausible. For the eq.(\ref{rational_roots}), one will have 
rational volume and hence central charge provided $\Delta\Lambda$ is
rational too. In any case, the exact computation of the Chern 
numbers might give us the exact result.  

In order to have a smooth geometry
in 5-d, the parameters need to obey \cite{clpp} $\alpha,\beta >x_2$ with 
$x_3 >x_2 \ge x_1 > 0$, which according to \cite{clpp} imply $q\ge p>0$ and
$p+q>r>0$. It is interesting to note that for $x_1=x_2$ the volume of the 
5-d space vanishes. In order to avoid that we shall however take $x_2>x_1$, 
i.e. $\Lambda > 0$. 

It is given in \cite{clpp} that the  Ricci tensor of this compact manifold 
is given by $R_{\mu\nu}=4\lambda g_{\mu\nu}$. For $\lambda=1$, one can very 
easily conclude that the geometry written in (\ref{cy}) is indeed a Calabi-Yau.

\section{The solution}

The back reaction of D3 and D5 branes onto a space with $R^{3,1}\times M_6$ 
makes the geometry to take a warped product from \cite{ks}. The supergravity
solution when $M_6$ is $Y^{(p,q)}$ is derived in \cite{ehk} and for a 
deformation to this with one free parameter $s$ in \cite{ssp} and in \cite{z}
a nearby space of  $Y^{(p,q)}$ has been studied by solving the Lichenerowicz 
equation and finding the supergravity solution in that geometry. 

To find an 
${\cal N}=1$ supersymmetry preserving solution, we do not have to show 
explicitly that the complex 3-form field strength constructed out of 3-form 
 RR and NSNS fields i.e. $G_3=F_3-\f{i}{g_s}H_3$ is a (2,1) form for the 
geometry whose $M_6$  part is described by 
(\ref{cy}). It is known that if $G_3$ satisfy ISD condition 
then its enough for that solution will preserve the above mentioned supersymmetry.

The ansatz to the 10-dimensional geometry is
\be
\label{metric_10}
ds^2=h^{-1/2}ds^2_4+h^{1/2}\bigg( dr^2+r^2[
{e^{\theta}}^2+{e^{\tau}}^2+{e^x}^2+{ e^{+}}^2+{ e^{-}}^2]\bigg).
\ee
where the warp factor $h$ is a function of several coordinates and the form of it will be determined later and  the 1-forms are defined, with $\lambda=1$, as
\beqn
& &e^x=\f{\rho}{2\sqrt{\Delta_x}}dx;\quad e^{\theta}=\f{\rho}{\sqrt{\Delta_{\theta}}} d\theta; \quad e^{\tau}=d\tau+\sigma\nn \\& &
e^+=\f{\sqrt{\Delta_x}}{\rho}\bigg(\f{sin^2\theta}{\alpha}d\phi+\f{cos^2\theta}{\beta}d\psi\bigg)\nn \\ & &
e^-=\f{\sqrt{\Delta_{\theta}} sin2 \theta}{2\rho}\bigg( \f{\alpha-x}{\alpha}d\phi-\f{\beta-x}{\beta}d\psi\bigg).
\eeqn

The supergravity solution is derived by taking the 
dilaton  to be a constant and the axion is set to zero i.e. $dilaton=g_s$ 
and $C_0=0$, respectively.

The two form, $B_2$, NSNS potential is assumed to take the form
\be
B_2= g_s MK lnr~~ \omega,
\ee
where M and K are constants and are related to normalization of flux and integration constant. The form of $\omega$ is 
\be
\omega=\f{1}{\sqrt{\Delta_x\Delta_{\theta}} sin2\theta} \Bigg(e^x\w e^{\theta}-e^-\w e^+\Bigg).
\ee
It is easy to see that this two form object possess several interesting properties like: closed, anti-selfdual  with respect to $x,\theta, \pm$ coordinates 
and $\omega\w J_4=0$, where $J_4=(1/2) d\sigma=e^{\theta}\w e^--e^x\w e^+$.

The form of 3-form RR and NSNS field strengths,  which satisfy ISD condition 
for $G_3=F_3-\f{i}{g_s} H_3$, are 
\be
F_3=MK e^{\tau}\w \omega; \quad H_3=g_s MK \f{dr}{r}\w\omega.
\ee  

The ansatz to the 5-form field strength ${\tilde F}_5$ is 
\be
g_s {\tilde F}_5= dh^{-1}\w dx^0\w\ldots\w dx^3+\star_{10}\bigg[ dh^{-1}\w dx^0\w\ldots\w dx^3 \bigg].
\ee
The  Bianchi identity 
associated to  3-form fields are  satisfied and are very easy to check
and the equation of motion associated to the 3-form field strength  do not
give us the form of the warp factor. The self duality on the five form field 
strength is automatic. In order to know 
the dependence of the warp factor on the coordinates one need to solve for 
  either the equation of ${\tilde F}_5$ or the equation of motion of
the metric  and we end up the following equation by assuming that the 
warp factor depends on $r,x,\theta$ coordinates i.e. $h=h(r,x,\theta)$.

\beqn
\label{h}
& &r^5 h^{''}_r+5r^4h^{'}_r+\f{4r^3}{\rho^2} (h^{''}_x\Delta_x+h^{'}_x\Delta^{'}_x)+\f{r^3}{\rho^2} (h^{''}_{\theta}\Delta_{\theta}+h^{'}_{\theta}\Delta^{'}_{\theta})=\nn \\& &
=-\f{2 g^2_s M^2 K^2}{r\Delta_x\Delta_{\theta}}-\f{r^3}{\rho^2} cot2\theta h^{'}_{\theta} \Delta_{\theta},
\eeqn
where $h^{'}_r$ denotes first partial derivative of h with respect to r and 
similarly for others i.e. prime denotes partial derivative and the number of 
times it appear says how many times derivative is being taken and the subscript denote coordinate with respect to which its being evaluated. 
Equation (\ref{h}) can be brought to a form where the warp factor can be
taken as 
\be 
h(r,x,\theta)=\f{h_1(r)+h_2(x,\theta)}{r^4},
\ee
and then using the separation of variable technique we find the solution to 
$h_1(r)$ as
\be
h_1(r)=-\f{A}{4}\bigg[ ln (r/r_0)-(C_2/A) r^4\bigg]; 
\quad ~~{\rm with }~~ln~r_0=4 C_3/A,
\ee
where $C_2,C_3$s are constants of integration and A is the constant that is 
used in the separation of variable of technique. $C_1\equiv 2g^2_s M^2K^2$. 

The equation for $h_2(x,\theta))$ 
is
\beqn
& &A\Delta_{\theta}+\bigg(\f{C_1}{\Delta_x}-Ax \bigg)-\f{C_1 x}{\Delta_x \Delta_{\theta}}+4(h^{''}_{2x}\Delta_x+h^{'}_{2x}\Delta^{'}_x)+\nn \\& &
(h^{''}_{2\theta}\Delta_{\theta}+h^{'}_{2\theta}\Delta^{'}_{\theta}+\Delta_{\theta} h^{'}_{2\theta} cot2 \theta)=0.
\eeqn

However, we have not managed to solve this equation.

\section{2-cycles}

In this section we are not going to follow the approach of \cite{ms} to find 
various supersymmetric 3-cycles. Rather, what we shall do is to find under
which situation the integration of  $B_2$ over  $S^2$
becomes non-zero, the purpose of evaluating these fluxes would be to 
make   connection with the field theoretic variables via  
AdS/CFT correspondence \cite{ks}, keeping 
in mind that there are degeneration points in the
5-d SE space \cite{clpp} and for the presence of 3-cycle follows from the 
argument that in a Calabi-Yau there exists a no where vanishing holomorphic 
3-form. The degeneration point is defined as a point in the 
space of coordinates where the length of the
Killing vector vanishes. For $L^{(p,q,r)}$ the degeneration points are at
$\theta=0,\pi/2$ and at $x=x_1,x_2$, the two lowest roots of $\Delta_x=0$.

Even though it seems that $B_2$ is diverging at these points but
we suggest to evaluate the integral of these form fields cautiously. 

The form of the  $B_2$  field is 
   
\be
B_2=\f{g_sMK ln~ r}{sin2\theta\sqrt{\Delta_x\Delta_{\theta}}}\Bigg[ \f{\rho^2}{2\sqrt{\Delta_x\Delta_{\theta}}}dx\w d\theta+\f{\sqrt{\Delta_x\Delta_{\theta}}}{{2\alpha\beta}} sin2\theta d\psi\w d\phi \Bigg].
\ee

Now, by doing a gauge transformation we can
gauge way the first piece as locally we can write 
down this term as an exact form, i.e. $d\Lambda$, and left with the second 
term. It is interesting to note 
that the integral of $B_2$ for constant $x$ and $\theta$ for the second term 
becomes a simple number. More importantly, to  fix the coordinates 
$x,\theta$ to
somevalue other than the value that it takes at the degeneration point and for 
these choice one can show that the volume of the 2-cycle to be non-zero.
So, the 2-cycle that we define is by fixing 
$x,\theta$ such that $\theta\ne 0,\pi/2$ and $x\ne x_1,x_2$ 
then integrating over $\psi,\phi$, i.e. 
\be
S^2=\int d\psi\w d\phi~~~~~~{\rm for ~~constant}~~~~~ (x, \theta)~~ {\rm with}
~~\theta\ne 0,\pi/2~~ {\rm and}~~  x\ne x_1,x_2.
\ee
The volume of this 2-cycle comes out to be 
\be
Vol_2=(2\pi)^2 \f{sin2\theta~\sqrt{\Delta_x\Delta_{\theta}}}{\alpha\beta}
\ee

We know that the Calabi-Yau constructed from this  SE spaces admit the 
topology of $S^2\times S^3$ \cite{ms1} and to justify the presence of the 
non-contractible 3-cycles,   we know the CY is supposed to 
admit a no where vanishing 
 holomorphic (3,0) form means we are guaranteed to have a 3-cycle \footnote{We 
would like to thank Chris Pope for letting us know that this SE spaces
do admit the above mentioned topology.}. \\

Note added: As we were finishing this paper there appeared  \cite{ms1} with 
which there are some overlaps of this work.

\section{Acknowledgment}
SSP would like to thank O. Aharony  for useful discussions and 
 C. Pope for a useful correspondence. The financial help
from Feinberg
graduate school  is gratefully acknowledged.

\end{document}